\documentclass[twocolumn,amssymb,amsmath,nobibnotes, aps, pra]{revtex4-1}
\usepackage{graphicx}
\usepackage{dcolumn}
\usepackage{bm}
\usepackage{amssymb,hyperref}
\usepackage{epstopdf}
\usepackage{color}
\usepackage{dsfont}
\usepackage{float}

\begin{document}

\title{Anti-chiral edge states in an exciton polariton strip}

\author{S. Mandal}\email{subhaska001@e.ntu.edu.sg}
\author{R. Ge}
\author{T. C. H. Liew}\email{tchliew@gmail.com}

\affiliation{Division of Physics and Applied Physics, School
of Physical and Mathematical Sciences, Nanyang Technological
University, Singapore 637371, Singapore}


\begin{abstract}
We present a scheme to obtain antichiral edge states in an exciton-polariton honeycomb lattice with strip geometry, where the modes corresponding to both edges propagate in the same direction. Under resonant pumping the effect of a polariton condensate with nonzero velocity in one linear polarization is predicted to tilt the dispersion of polaritons in the other, which results in an energy shift between two Dirac cones and the otherwise flat edge states become tilted. Our simulations show that due to the spatial separation from the bulk modes the edge modes are robust against disorder. 
\end{abstract}

\maketitle

\section{Introduction}
Topological phases of matter have received tremendous attention in the scientific community for the last few decades due to the presence of robust edge states in between gapped bulk states, which make the edges of finite samples conducting even with an insulating bulk. Topological electronic systems with broken time reversal (TR) symmetry \cite{Laughlin_1981,Haldane_1988} are characterized by counter propagating chiral edge states whereas TR invariant topological systems contain a pair of counter propagating helical edge states at each edge corresponding to different spins states and known as the quantum spin Hall effect \cite{Kane_2005}. The application of topology in bosonic systems has resulted in creation of one-way transport of photons \cite{Lu_2014, Haldane_2008, Wang_2009, Koch_2010, Rechtsman_2013, Hafezi_2013}, excitons \cite{Zhou_2014, Bardyn_2015} and  exciton-polaritons \cite{Bardyn_2015, Karzig_2015, Nalitov_2015, Bardyn_2016, Sigurdsson_2017, Whittaker_2018, Banerjee_2018, Ge_2018, Klembt_2018}. Very recently antichiral edge states have been proposed in the modified Haldane model, where modes corresponding to both the edges of a strip flow in the same direction, compensated by the conducting bulk modes that propagate in the opposite direction \cite{Colomes_2018}. In this paper we present an exciton polariton based scheme to obtain antichiral edge states in a photonic system.

Exciton polaritons are  hybrid particles of excitons and cavity photons. Due to their excitonic fraction, polaritons are strongly interacting, which enables them to form Bose-Einstein condensates and to have quantum fluid nature \cite{Carusotto_2013}. Initial proposals for creating topological polaritons lied in the linear regime, not exploiting interactions between themselves, while interaction with a hot exciton reservoir was shown to provide a gain mechanism for topological polariton lasing \cite{Klembt_2018}. Theoretically, nonlinearity may lead to interaction induced topology \cite{Bardyn_2016}, solitons \cite{Kartashov_2016,Gulevich_2017}, and bistable topological polaritons  \cite{Kartashov_2017}. Here, we make use of the strong nonlinear polariton-polariton interaction in a honeycomb lattice with zigzag edges to realize anti-chiral edge states. Since the total number of right moving  and total number of left moving modes must be equal, it is impossible to realize the antichiral edge states in a gapped band structure. But in a gapless band structure, where the counter propagating modes are provided by bulk modes, antichiral edge states become possible \cite{Colomes_2018}. Due to the spatial separation of the edge modes from the bulk modes backscattering is significantly suppressed in these systems similar to the case in topological insulators.
\begin{figure}[b]
\includegraphics[width=0.45\textwidth]{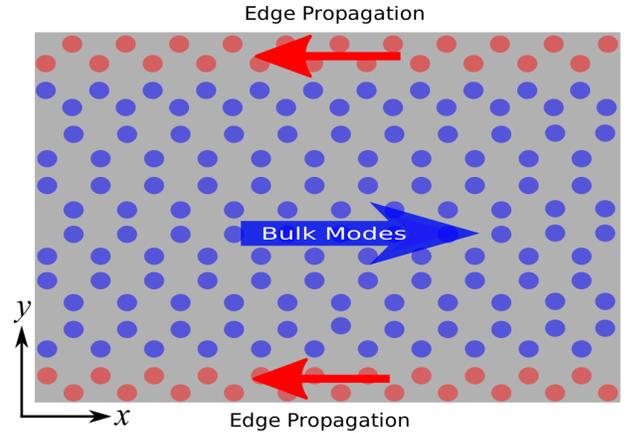}
\caption{Schematic view of the system under consideration. A graphene like polariton strip with zigzag edges, which can be fabricated with different technologies \cite{Kim_2011, Mendez_2013, Jacqmin_2014}, is subjected to an $x$ linearly polarized optical field. $y$ linearly polarized polaritons propagate at both the edges in the same direction compensated by counter propagating polaritons through the bulk. Since the edge and bulk modes are spatially separated, the edge states are robust against scattering with disorder.}
\label{fig:fig1}
\end{figure}
\section{Theoretical Scheme}
We consider polaritons under resonant excitation in the $x$ and $y$ linearly polarized basis described by the driven dissipative Gross-Pitaevskii equation \cite{Shelykh_2010}
\begin{align}\label{x_y_polarized}
i\hbar\frac{\partial \phi_{x,y}}{\partial t} &= \bigg[\varepsilon_{x,y}-\frac{\hbar^2\nabla^2}{2m}-i\frac{\Gamma_{x,y}}{2}+V(\mathbf{x})+U_0\Big(|\phi_x|^2\nonumber\\
&+|\phi_y|^2\Big)\bigg]\phi_{x,y}-U_1\Big(|\phi_{x,y}|^2\phi_{x,y}+\phi_{y,x}^2\phi^*_{x,y}\Big)\nonumber\\
&+F_{x,y}(\mathbf{x})e^{-i\omega_0t}
\end{align}
where $\varepsilon_x$ and $\varepsilon_y$ are the energies of the $x$ and $y$ polarized polaritons with lifetime $\Gamma_x$ and $\Gamma_y$ respectively, $\nabla^2$ is the Laplacian operator, $m$ is the effective mass of the polaritons and $V(x)$ is the potential. The nonlinear polariton-polariton interaction constants can be expressed in terms of those in the spinor basis by $U_0=\alpha_1$ and $U_1=(\alpha_1-\alpha_2)/2$ \cite{Shelykh_2010}. It is well known that polaritons with the same spin interact repulsively making $\alpha_1$ positive, whereas  polaritons with opposite spin interact attractively making $\alpha_2$ negative and typically $\alpha_1\geq |\alpha_2|$ \cite{Krizhanovskii_2006,Leyder_2007}. $F_{x,y}$ is a polarization dependent resonant incident optical field with frequency $\omega_0$ and $\phi_{x,y}$ are the polarization dependent polariton wave functions.

In what follows, we will excite the $x$ polarized polaritons using the resonant excitation for which $F_y(\mathbf{x})=0$. In such a limit the population of $x$ polarized polaritons is much stronger than that of $y$ polarized polaritons. Nevertheless, we are interested in the dispersion of $y$ polarized polaritons, which can still be weakly populated by fluctuations or additional pulses. We can further simplify the problem by choosing the strength of the polariton polariton interaction strengths to be equal, $U_0=U_1$, which has been realized in the experiments \cite{Vladimirova_2010,Takemura_2014}. In this limit terms $|\phi_{x,y}|^2\phi_{x,y}$ vanish from Eq.~\ref{x_y_polarized}. Given that $y$ polarized polaritons are weakly populated, we neglect second order terms in  $\phi_{y}$ (This assumption is particularly accurate as it is readily seen that $\phi_{y}=0$ is a solution to Eq.~\ref{x_y_polarized} when $F_y(\mathbf{x})=0$ and stability analysis reveals that it is a stable solution). After taking all the above mentioned conditions into account Eq.~\ref{x_y_polarized} becomes: 
\begin{align}\label{x_y_polarized_2}
i\hbar\frac{\partial \phi_{x}}{\partial t} = & \left[\varepsilon_{x}-\frac{\hbar^2\nabla^2}{2m}-i\frac{\Gamma_{x}}{2}+V(\mathbf{x})\right]\phi_{x}+F_{x}(\mathbf{x})e^{-i\omega_0t}\nonumber\\
i\hbar\frac{\partial \phi_{y}}{\partial t} = & \left[\varepsilon_{y}-\frac{\hbar^2\nabla^2}{2m}-i\frac{\Gamma_{y}}{2}+V(\mathbf{x})+U_0|\phi_{x}|^2\right]\phi_{y}\nonumber\\
&-U_0\phi_{x}^2\phi^*_{y}\nonumber\\
\end{align}
We choose,
\begin{align}\label{Potential}
V(\mathbf{x})=& V_0\Bigg[\cos\left(\frac{4\pi y}{\sqrt{3}a}\right)+\cos\left(-\frac{2\pi x}{a}-\frac{2\pi y}{\sqrt{3}a}\right)\nonumber\\
& +\cos\left(\frac{2\pi x}{a}-\frac{2\pi y}{\sqrt{3}a}\right)\Bigg]
\end{align}
which corresponds to a honeycomb lattice with periodicity $a$ in the $x$ direction and  zig-zag edges if a cut is made making the system finite in the $y$ direction. Although we consider here a smooth potential, similar results can be obtained with a square well type potential at each lattice site. To simplify our analysis we focus on the slowly varying field $\psi_{x,y}=\phi_{x,y}e^{i\omega_0 t}$ and make the following substitutions to move to a dimensionless unit system $t\rightarrow (2ma^2/\hbar) t$, $x\rightarrow ax$, $y\rightarrow ay$, $\varepsilon_{x,y}\rightarrow(\hbar^2/2ma^2)\varepsilon_{x,y}$ $V_0\rightarrow(\hbar^2/2ma^2)V_0$, $\Gamma_{x,y}\rightarrow (\hbar^2/2ma^2)\Gamma_{x,y}$, $\omega_0\rightarrow (\hbar/2ma^2)\omega_0$, $\psi_{x,y}\rightarrow(\sqrt{\hbar^2/2ma^2})\psi_{x,y}/\sqrt{U_0}$ and $F_x\rightarrow F_x\sqrt{U_0}(2ma^2/\hbar^2)^{\frac{3}{2}}$. In the dimensionless units, Eq.~\ref{x_y_polarized_2} takes the form
\begin{align}
i\frac{\partial \psi_{x}}{\partial t} = & \left[\varepsilon_{x}-\omega_0-\nabla^2-i\frac{\Gamma_{x}}{2}+V(\mathbf{x})\right]\psi_{x}+F_{x}(\mathbf{x})\label{x_y_polarized_3}\\ 
i\frac{\partial \psi_{y}}{\partial t} = & \left[\varepsilon_{y}-\omega_0-\nabla^2-i\frac{\Gamma_{y}}{2}+V(\mathbf{x})+|\psi_{x}|^2\right]\psi_{y}-\psi_{x}^2\psi^*_{y}\label{x_y_polarized_4}
\end{align}
\section{Band Structure}
As mentioned earlier our aim is to study the dispersion of the $y$ polarized polaritons, which can be calculated by substituting $\psi_{y}=u(\mathbf{x})e^{i\omega t}+v^*(\mathbf{x})e^{-i\omega^* t}$ in Eq.~\ref{x_y_polarized_4}, where $u(\mathbf{x})$ and $v^*(\mathbf{x})$ are spatial functions to be found and $\omega$ is the frequency, which is kept complex in order to capture potential instabilities \cite{Carusotto_2004} and we continue to work in the limit where the population of $\psi_y$ is small.
\begin{figure}[b]
\includegraphics[width=0.5\textwidth]{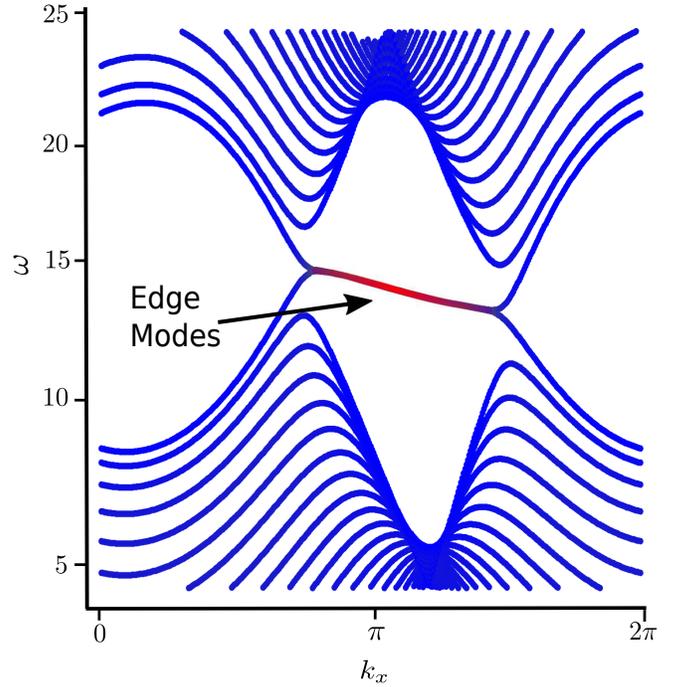}
\caption{Dispersion of the $y$ polarized polaritons  calculated from Eq.~\ref{dispersion_matrix} corresponding to the stable states. All the states marked in blue correspond to the bulk states and states marked in red correspond to the edge states. Parameters: $\omega_0=15$, $k_0=\pi$, $\varepsilon_x=0$, $\varepsilon_y=A_0^2$, $A_0=3.6$, $\Gamma_x=\Gamma_y=0.23$, $V_0=17.7$, $\zeta=0.1$. }
\label{fig:fig2}
\end{figure}
 Upon substitution, Eq.~\ref{x_y_polarized_4} becomes two coupled equations in $u(\mathbf{x})$ and $v(\mathbf{x})$ which can be expressed in the following matrix form
\begin{equation}\label{dispersion_matrix}
\begin{bmatrix}
    \omega^{\prime}(\mathbf{x})      & -\psi^2_{x} \\
     \psi^{*2}_{x}      & -\omega^{\prime*}(\mathbf{x})\\
\end{bmatrix}
\begin{bmatrix}
   u(\mathbf{x}) \\
    v(\mathbf{x})\\
\end{bmatrix}
=
\omega\begin{bmatrix}
   u(\mathbf{x}) \\
    v(\mathbf{x})\\
\end{bmatrix}
\end{equation}
where $\omega^{\prime}(\mathbf{x})=\varepsilon_{y}-\omega_0-\nabla^2+|\psi_{x}|^2-i{\Gamma_{y}}/{2}+V(\mathbf{x})$. It is  well known that graphene with zigzag edges shows flat edge dispersion with zero group velocity. Our aim is to make the dispersion asymmetric such that the edge states gain  non zero group velocity. It was previously shown that a resonant excitation with non-zero in-plane wavevector  can tilt the dispersion of excitations \cite{Carusotto_2004}.  The first and principle ingredient of our scheme consists of a stationary field $\psi_x$ with nonzero wave vector which tilts the dispersion of $\psi_y$. We choose $\psi_x=A(y)e^{ik_0x}$, where $\psi_x$  vanishes at the boundaries $[-y_L,y_L]$ . To take the boundary conditions into account we take (for $|y|<y_L$)
\begin{equation}\label{A_y}
A(y)=A_0\left[1-e^{-{(y-y_L)^2}/{2\zeta^2}}\right]\times \left[1-e^{-{(y+y_L)^2}/{2\zeta^2}}\right]
\end{equation}
This form of $\psi_x$ can be easily imprinted by a specific choice of the driving field, which corresponds to a transverse electric polarized laser excitation with non-zero angle of incidence,
\begin{equation}\label{pump_form}
F_{x}(\mathbf{x})= \left[-\varepsilon_{x}+\omega_0+\nabla^2+i\frac{\Gamma_{x}}{2}-V(\mathbf{x})\right]A(y)e^{ik_0x},
\end{equation}
\begin{figure}[b]
\includegraphics[width=0.5\textwidth]{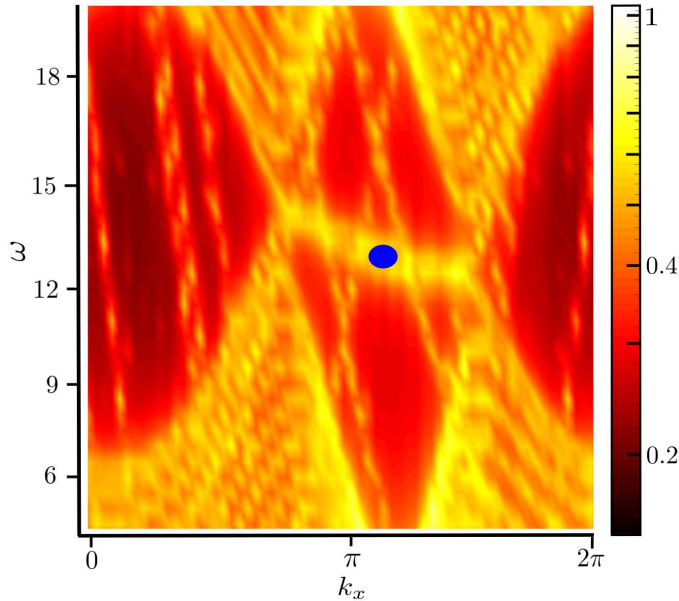}
\caption{Dispersion of the $y$ polarized polaritons calculated from Eq.~\ref{x_y_polarized_3} - \ref{x_y_polarized_4} in the presence of white noise which corresponds to the photoluminescence spectrum obtainable in experiments. The blue dot corresponds to the state which is excited to probe the pulse propagation.}
\label{fig:fig3}
\end{figure}
which can be found by substituting the form of $\psi_x$  in Eq.~\ref{x_y_polarized_3} and solving for $F_x(\mathbf{x})$. To keep Eq.~\ref{dispersion_matrix} periodic in the x direction the only nontrivial permitted values of $k_0$ are $(2n+1)\pi$, where $n$ can be zero or any integer. Under this condition, we can apply Bloch theory on Eq.~\ref{dispersion_matrix} to write down 
\begin{equation} \label{uandv}
 u(\mathbf{x})=e^{ik_x x}\tilde{u}_{k_x}(\mathbf{x}), ~~v(\mathbf{x})=e^{i k_x x}\tilde{v}_{k_x}(\mathbf{x}).
\end{equation}
where $\tilde{u}_{k_x}(\mathbf{x})$ and $\tilde{v}_{k_x}(\mathbf{x})$ are the Bloch wave functions of the  fluctuations with wave vector $k_x$. Substituting Eq.~\ref{uandv} into Eq.~\ref{dispersion_matrix} we get the following equations which are used to calculate the bandstructure in Fig.~\ref{fig:fig2}
\begin{align}\label{big_dispersion}
&\left[(k_x+G)^2-\frac{\partial^2}{\partial y^2}+\varepsilon_y+|A(y)|^2-i\frac{\Gamma_y}{2}-\omega_0-\omega\right]\tilde{u}(G,y)\nonumber\\
&+\left[\sum_{G^\prime}\tilde{V}(G-G^\prime,y)\tilde{u}(G^\prime,y)-\tilde{\psi}_x^{2^*}(G^\prime-G,y)\tilde{v}(G^\prime,y)\right]=0\nonumber\\
&\left[-(k_x+G)^2+\frac{\partial^2}{\partial y^2}-\varepsilon_y-|A(y)|^2-i\frac{\Gamma_y}{2}+\omega_0-\omega\right]\tilde{v}(G,y)\nonumber\\
&-\left[\sum_{G^\prime}\tilde{V}(G-G^\prime,y)\tilde{v}(G^\prime,y)-\tilde{\psi}_x^{2}(G-G^\prime,y)\tilde{u}(G^\prime,y)\right]=0
\end{align} 
where $G=2n\pi$ is the reciprocal wave vector with $n$ can be zero or any integer and using Fourier transformation the following expressions can be found
\begin{align}
\tilde{u}_{k_x}(\mathbf{x})&=\sum_{G}\tilde{u}(G,y)e^{iGx}, ~~\tilde{u}(G,y)=\displaystyle\int_{-1/2}^{1/2}\tilde{u}_{k_x}(\mathbf{x})e^{-iGx},\nonumber\\
\tilde{v}_{k_x}(\mathbf{x})&=\sum_{G}\tilde{v}(G,y)e^{iGx}, ~~\tilde{v}(G,y)=\displaystyle\int_{-1/2}^{1/2}\tilde{v}_{k_x}(\mathbf{x})e^{-iGx},\nonumber\\
V(\mathbf{x})&=\sum_{G}\tilde{V}(G,y)e^{iGx}, ~~\tilde{V}(G,y)=\displaystyle\int_{-1/2}^{1/2}V(\mathbf{x})e^{-iGx},\nonumber\\
\psi_x^2(\mathbf{x})&=\sum_{G}\tilde{\psi}_x^{2}(G,y)e^{iGx}, ~~\tilde{\psi}_x^{2}(G,y)=\int_{-1/2}^{1/2}\psi_x^2(\mathbf{x})e^{-iGx}.
\end{align}
Since it is a non-hermitian system the spectrum will be complex and only the real part of the spectrum near the Dirac points is plotted in Fig~\ref{fig:fig2}. The spectrum corresponding to $\omega<0$ is the image of that corresponding to $\omega>0$ under transformation $k\rightarrow 2k_0-k$ and $\omega\rightarrow2\omega_0-\omega$ \cite{Carusotto_2004}.  In physical units, we have found that the energy shift between the two Dirac points can be around $0.1$ meV, taking $m=7\times 10^{-5}m_e$, where $m_e$ is the free electron mass and $a=2.2~\mu m$. This exceeds the decay rates in modern samples where polaritons with long lifetime are obtained \cite{Sun_2017}. The required nonlinear shift due to the polariton-polariton interaction to obeserve the effect is about $1.45$ mev which is within experimental limits \cite{YSun_2017,Wertz_2011,Anton_2013}. Another requirement of this scheme is that the chosen energy and wave vector of the pump needs to be  within the parabolic region of the lower polariton branch so that parametric instabilities are avoided \cite{Carusotto_2013}. Since these edge states are spatially separated from the bulk states the backscattering of the edge states should be significantly suppressed \cite{Colomes_2018}. Nevertheless there  is some overlap between the bulk and edge states and the suppression of backscattering is not expected to be as strong as in the topologically protected case.
\section{Results and Discussions}
\begin{figure}[H]
\centering
\includegraphics[width=0.5\textwidth]{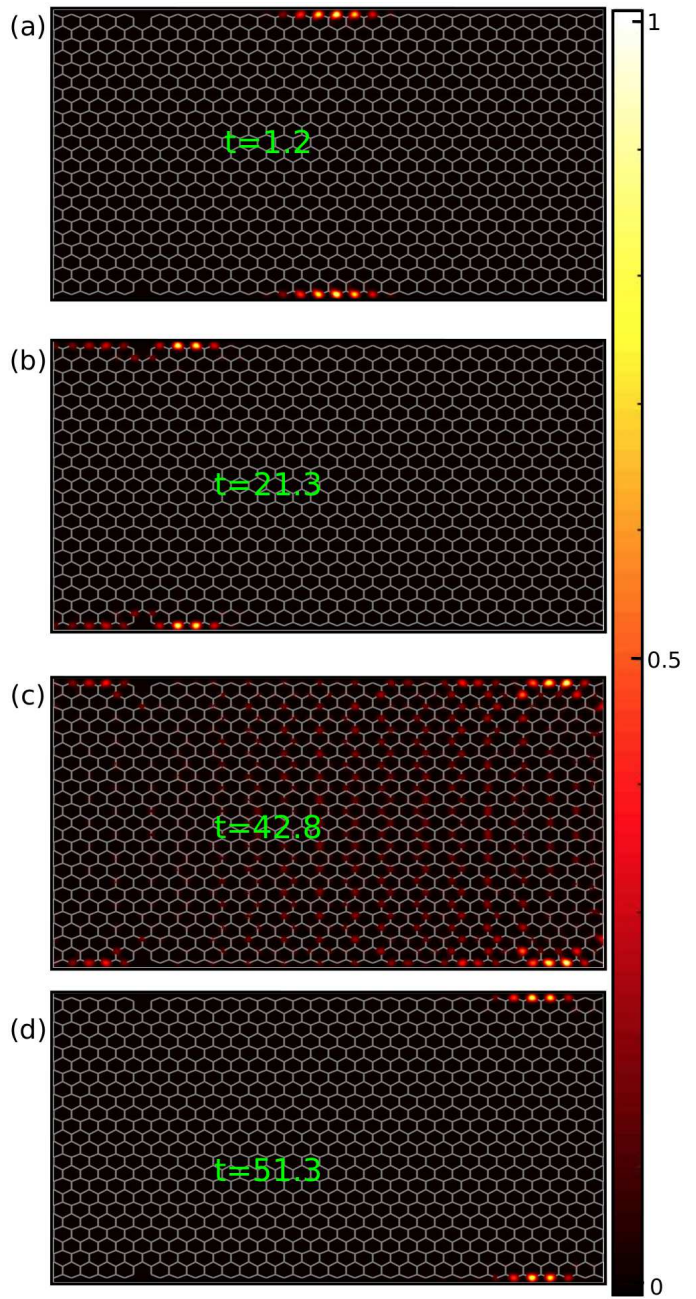}
\caption{Propagation of the polaritons in a graphene zigzag strip consisting of 32$\times$23 unit cells with both the edges slightly deformed. Polaritons propagate through both the edges in the same direction and go around the defects without being backscattered (a - b). Once the polaritons reach the left end they couple to the counter propagating bulk modes and the intensity is  transferred back to the right end (c). When the polaritons reach to the right end through the bulk they again couple to the edge modes and start to propagate through the edges robustly until they again reach the left end (d). The energy and wave vector of the pulse correspond to the blue dot shown in Fig~\ref{fig:fig3}.  Due to the finite lifetime of the polaritons the intensity decreases with time, which is compensated by rescaling of the intensity at each time step. Parameters: $F_0=2$, $\sigma=2$, $\tau=4$, $t_0=0$.}
\label{fig:fig4}
\end{figure}

\begin{figure}[t]
\includegraphics[width=0.5\textwidth]{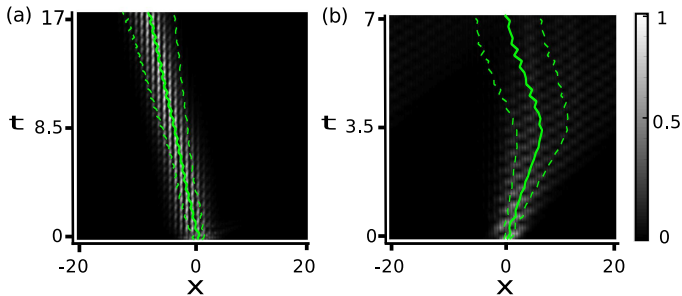}
\caption{The propagation of an edge state (a) and a bulk state (b) in presence of disorder. Due to the absence of backscattering the edge state does not spread too much whereas the bulk state spreads indicating backscattering. The green line represents the mean position of the pulse and the dashed green lines represent the square root of variance.}
\label{fig:fig5}
\end{figure}

To verify our claim we evolve Eqs.~\ref{x_y_polarized_3} , \ref{x_y_polarized_4} in time in the presence of an additional stochastic complex Langevin noise term (chosen as a white noise in space and time). Once $\psi_x$ reaches its steady state one can obtain the dispersion corresponding to $\psi_y$ as shown in Fig~\ref{fig:fig3}. This corresponds to the photoluminescence spectrum that could be obtained experimentally. To have consistency, all the parameters are kept the same as the ones in Fig~\ref{fig:fig2}. To illustrate the propagation and robustness of the edge states we introduced an impurity by slightly deforming the potential at the edges and then excited  both the edges using a $y$ polarized coherent pulse of the form
\begin{align}
F_y=&F_0\Big[\exp \{-[(x-x_0)^2+(y-y_L)^2]/2\sigma^2 \} \nonumber\\
&+\exp \{-[(x-x_0)^2+(y+y_L)^2]/2\sigma^2 \}\Big]\nonumber\\
&\times \exp[-(t-t_0)^2/\tau^2 ]\exp[i(k_p x-\omega_p t)]
\end{align}
where $F_0$,  $\tau$  and $\omega_p$ are the amplitude, duration and frequency of the pulses respectively centered at $(x_0,y_L)$ and $(x_0,-y_L)$ having wave vector $k_p$ with width $\sigma$. The values of $k_p$ and $\omega_p$ correspond to the blue dot shown in Fig.~\ref{fig:fig3}. Different stages of the pulse propagation are shown in Fig~\ref{fig:fig4}.  As expected both the pulses propagate in the same direction along the edges of the sample. Even in the presence of the impurities the pulses do not backscatter but flow around the impurities which prove the robustness of the edge states. However, once the pulses reach the left end of the sample they have no where to go but to couple to the counter propagating bulk modes (which propagate from left to right) and the intensity can be seen to be transfered towards the right end of the sample through the bulk. Bulk modes at the right edge couple to the edge states and the intensity is  transfered back to the edges where they again start to propagate from right to left. To compare the robustness of an edge state to  a bulk state we add disorder in the system by adding a Gaussian correlated disorder potential with the honeycomb lattice potential $V(\mathbf{x})$  and excite an edge state and a bulk state with a coherent Gaussian pulse. In Fig~\ref{fig:fig5} we plot the intensity distribution along the propagating axis with time by summing over the other axis. The edge state can be seen to propagate without being backscattered (Fig.~\ref{fig:fig5}(a)), whereas the bulk state spreads over the $x$ axis with time indicating the presence of backscattering (Fig.~\ref{fig:fig5}(b)). The green line represents the mean position of the pulse and the dashed green lines represent the square root of variance around the mean position. We have checked that the unidirectional propagation of the antichiral edge states is unhampered for disorder strength less than $30~\mu eV$,
which is consistent with the experimental observation \cite{Baboux_2016}. We have checked that unlike \cite{Colomes_2018} this system does not leave behind a half topological charge. Although the propagation length considered in Fig.~\ref{fig:fig5} is long enough to show a clear advantage of anti-chiral edge states compared to  bulk states, it is not enough to show the possible coupling of antichiral edge states to bulk states at the same energy. This can potentially be caused by disorder in the system, so we stress that we are not claiming that antichiral edge states are as robust as compared to the chiral edge states in topological systems. These anti-chiral edge states can nevertheless be interesting in polaritonic systems for building polariton wires where information could be efficiently transfered  along the edges due to the absence of backscattering. It is typically claimed that polariton gap solitons in nanowires and chiral edge states in lattices can be used in information processing polaritonic devices. However, criteria for photonic information processing were established long ago \cite{Keyes_1985}, where it was pointed out that it is essential for any feedback to be suppressed in the system. Polariton gap solitons can propagate in both forward and backward directions, while chiral edge states always come in forward and backward propagating pairs. In contrast antichiral edge states propagate in the same direction at both edges and thus correspond to a feedback suppression mechanism which make them an excellent candidate for information processing photonic devices\cite{Sanvitto_2013}.
\section{Conclusion}
In summary, we have discussed a theoretical scheme to obtain antichiral edge states in a photonic system, making use of the polarization dependent interactions of polariton condensates, where the motion of an $x$ linearly polarized polariton condensate tilts the dispersion of  $y$ linearly polarized polaritons. A non zero velocity can be easily induced in a polariton condensate by the choice of a coherent optical incident field with non zero in plane wave vector. The two key ingredients in our scheme involve an  exciton polariton honeycomb lattice with zigzag edges and a non zero velocity of the $x$ polarized polariton condensates, which then shifts the two Dirac points of the $y$ linearly polarized polaritons in energy resulting in antichiral edge states where the modes corresponding to both the edges propagate in the same direction.  Due to the spatial separation of the edge and bulk modes backscattering is suppressed even in the presence of disorder. 
\section{Acknowledgements}
The work was supported by the Ministry of Education (Singapore), grants 2017-T2-1-001.


\begin{thebibliography}{99}
\bibitem{Laughlin_1981}
R. B. Laughlin, Phys. Rev. B {\bf 23}, 5632(R)  (1981).

\bibitem{Haldane_1988}
F. D. M. Haldane, Phys. Rev. Lett {\bf 61}, 2015  (1988).

\bibitem{Kane_2005}
C. L. Kane, and E. J. Mele, Phys. Rev. Lett {\bf 95}, 226801  (2005).

\bibitem{Lu_2014}
 L. Lu, J. D. Joannopoulos, and M. Soljacic, Nat. Photonics {\bf 8}, 821 (2014).
 
 \bibitem{Haldane_2008}
 F. D. M. Haldane and S. Raghu, Phys. Rev. Lett. {\bf 100}, 013904 (2008).

 \bibitem{Wang_2009}
 Z. Wang, Y. Chong, J. D. Joannopoulos, and M. Soljacic, Nature (London) {\bf 461}, 772 (2009).

 \bibitem{Koch_2010}
J. Koch, A. A. Houck, K. L. Hur, and S. M. Girvin, Phys. Rev. A {\bf 82}, 043811 (2010).

 \bibitem{Rechtsman_2013}
 M. C. Rechtsman, J. M. Zeuner, Y. Plotnik, Y. Lumer, D. Podolsky, F. Dreisow, S. Nolte, M. Segev, and A. Szameit,  Nature (London) {\bf 496}, 196 (2013).

 \bibitem{Hafezi_2013}
 M. Hafezi, S. Mittal, J. Fan, A. Migdall, and J. M. Taylor, Nat. Photonics {\bf 7}, 1001 (2013).

\bibitem{Zhou_2014}
J. Yuen-Zhou, S. S. Saikin, N. Y. Yao, and A. Aspuru-Guzik, Nat. Mater. {\bf 13}, 1026 (2014).


\bibitem{Karzig_2015}
T. Karzig, C.-E. Bardyn, N. Lindner, and G. Refael, Phys. Rev. X {\bf 5}, 031001 (2015).


\bibitem{Bardyn_2015}
 C. E. Bardyn, T. Karzig, G. Refael, and T. C. H. Liew, Phys. Rev. B  {\bf 91}, 161413(R) (2015).



\bibitem{Nalitov_2015}
 A. V. Nalitov, D. D. Solnyshkov, and G. Malpuech, Phys. Rev. Lett. {\bf 114}, 116401 (2015).

\bibitem{Bardyn_2016}
C. E. Bardyn, T. Karzig, G. Refael, and T. C. H. Liew, Phys. Rev. B {\bf 93}, 020502(R) (2016).

\bibitem{Sigurdsson_2017}
H. Sigurdsson, G. Li, and T. C. H. Liew, Phys. Rev. B {\bf 96}, 115453 (2017).


\bibitem{Whittaker_2018}
C. E. Whittaker, E. Cancellieri, P. M. Walker, D. R. Gulevich, H. Schomerus, D. Vaitiekus, B. Royall, D. M. Whittaker, E. Clarke, I. V. Iorsh, I. A. Shelykh, M. S. Skolnick, and D. N. Krizhanovskii, Phys. Rev. Lett. {\bf 120}, 097401 (2018).

\bibitem{Banerjee_2018}
R. Banerjee, T. C. H. Liew, O. Kyriienko, Phys. Rev. B {\bf 98}, 075412 (2018).

\bibitem{Ge_2018}
R. Ge, W. Broer, and T. C. H. Liew, Phys. Rev. B {\bf 97}, 195305 (2018).



\bibitem{Klembt_2018}
S. Klembt, T. H. Harder, O. A. Egorov, K. Winkler, R. Ge, M. A. Bandres, M. Emmerling, L. Worschech, T. C. H. Liew, M. Segev, C. Schneider, and S. Hofling, Nature {\bf 562}, 552 (2018). 

\bibitem{Colomes_2018}
E. Colomes, and M. Franz, Phys. Rev. Lett {\bf 120}, 086603 (2018).


\bibitem{Carusotto_2013}
I. Carusotto, and C. Ciuti, Rev. Mod. Phys.{\bf 85}, 299 (2013).

\bibitem{Kartashov_2016}
Y. V. Kartashov, and D. V. Skryabin, Optica {\bf 3}, 1228 (2016). 

\bibitem{Gulevich_2017}
D. R. Gulevich, D. Yudin, D. V. Skryabin, I. V. Iorsh, and I. A. Shelykh, Sci. Rep., {\bf 7}, 1780 (2017). 


\bibitem{Kartashov_2017}
Y. V. Kartashov, and D. V. Skryabin, Phys. Rev. Lett {\bf 119}, 253904 (2017).


\bibitem{Kim_2011}
N. Y. Kim, K. Kusudo, C. Wu, N. Masumoto, A. Loffler, S. Hofling, N. Kumada, L. Worschech, A. Forchel, and
Y. Yamamoto, Nat. Phys. {\bf 7}, 681 (2011).

\bibitem{Mendez_2013}
E. A. Cerda-Mendez, D. Sarkar, D. N. Krizhanovskii, S. S.Gavrilov, K. Biermann, M. S. Skolnick, and P. V. Santos,
Phys. Rev. Lett. {\bf 111}, 146401 (2013).

\bibitem{Jacqmin_2014}
 T. Jacqmin, I. Carusotto, I. Sagnes, M. Abbarchi, D. Solnyshkov, G. Malpuech, E. Galopin, A. Lemaitre, J.
Bloch, and A. Amo, Phys. Rev. Lett. {\bf 112}, 116402 (2014).
%



\bibitem{Shelykh_2010}
I. A. Shelykh, A. V. Kavokin, Y. G. Rubo, T. C. H. Liew, and G.
Malpuech, Semicond. Sci. Technol. {\bf 25}, 013001 (2010).

\bibitem{Krizhanovskii_2006}
D. N. Krizhanovskii, D. Sanvitto, I. A. Shelykh, M. M. Glazov, G. Malpuech, D. D. Solnyshkov, A. Kavokin, S. Ceccarelli, M. S. Skolnick, and J. S. Roberts, Phys. Rev. B {\bf 73}, 073303 (2006).

\bibitem{Leyder_2007}
C. Leyder, T. C. H. Liew, A. V. Kavokin, I. A. Shelykh, M. Romanelli, J. P. Karr, E. Giacobino, and A. Bramati, Phys. Rev. Lett. {\bf 99}, 196402 (2007).

\bibitem{Carusotto_2004}
I. Carusotto and C. Ciuti, Phys. Rev. Lett. {\bf 93}, 166401 (2004).

\bibitem{Vladimirova_2010}
M. Vladimirova, S. Cronenberger, D. Scalbert, K. V. Kavokin,
A. Miard, A. Lemaître, J. Bloch, D. Solnyshkov, G. Malpuech,
and A. V. Kavokin, Phys. Rev. B {\bf 82}, 075301 (2010).

\bibitem{Takemura_2014}
N. Takemura, S. Trebaol, M. Wouters, M. T. Portella-Oberli, and
B. Deveaud, Phys. Rev. B {\bf 90}, 195307 (2014).

\bibitem{Sun_2017}
Y. Sun, P. Wen, Y. Yoon, G. Liu, M. Steger, L. N. Pfeiffer, K. West, D. W. Snoke, and K. A. Nelson, Phys. Rev. Lett. {\bf 118}, 016602 (2017).


\bibitem{YSun_2017}
Y. Sun, Y. Yoon, M. Steger, G. Liu, L. N. Pfeiffer, K. West, D. W. Snoke, and K. A. Nelson, Nat. Phys. {\bf 13}, 870 (2017).


\bibitem{Wertz_2011}
E. Wertz, A. Amo, D. D. Solnyshkov, L. Ferrier, T. C. H. Liew, D. Sanvitto, P. Senellart, I. Sagnes, A. Lemaitre, A. V. Kavokin, G. Malpuech, and J. Bloch, Phys. Rev. Lett. {\bf 109}, 216404, (2012).

\bibitem{Anton_2013}
C. Anton, T. C. H. Liew, J. Cuadra, M. D. Martin, P. S. Eldridge, Z. Hatzopoulos, G. Stavrinidis, P. G. Savvidis, and L. Vina, Phys. Rev. B {\bf 88}, 245307  (2013).

\bibitem{Baboux_2016}
F. Baboux, L. Ge, T. Jacqmin, M. Biondi, E. Galopin, A. Lemaître, L. Le Gratiet, I. Sagnes, S. Schmidt, H. E. Tureci, A. Amo, and J. Bloch, Phys. Rev. Lett. {\bf 109}, 066402 (2016).

\bibitem{Keyes_1985}
R W Keyes, Science, 230, 138 (1985)

\bibitem{Sanvitto_2013}
D. Sanvitto, and S. Kena-Cohen, Nat. Mater. {\bf 15}, 1061 (2016).

\end{thebibliography}
\end{document}